\documentclass[%
 reprint,
 superscriptaddress,
 amsmath,amssymb,
 aps,
]{revtex4-2}

\usepackage{graphicx}
\usepackage{dcolumn}
\usepackage{bm}
\usepackage{xcolor}
\usepackage{algorithm}
\usepackage{algorithmicx}
\usepackage[noend]{algpseudocode}
\usepackage[normalem]{ulem}

\usepackage[utf8]{inputenc}
\usepackage[T1]{fontenc}
\usepackage{mathptmx}
\usepackage[normalem]{ulem}

\usepackage{tikz}
\usetikzlibrary{shapes.geometric, arrows}
\tikzstyle{startstop} = [rectangle, rounded corners, minimum width=3cm, minimum height=1cm,text centered, draw=black, fill=white!30]
\tikzstyle{arrow} = [thick,->,>=stealth]

\newcommand {\av}[1]{ \langle #1  \rangle }
\newcommand {\pder}[2]{\frac{\partial  #1}{\partial #2} }

\begin{document}

\preprint{APS/123-QED}

\title{Automated free energy calculation from atomistic simulations}

\author{Sarath Menon}
\email{sarath.menon@rub.de}
\affiliation{Interdisciplinary Centre for Advanced Materials Simulation, Ruhr-Universit{\"a}t Bochum, 44801 Bochum, Germany}

\author{Yury Lysogorskiy}
\affiliation{Interdisciplinary Centre for Advanced Materials Simulation, Ruhr-Universit{\"a}t Bochum, 44801 Bochum, Germany}

\author{Jutta Rogal}
\affiliation{Department of Chemistry, New York University, 10003 New York, United States}
\affiliation{Fachbereich Physik, Freie Universit{\"a}t Berlin, 14195 Berlin, Germany }

\author{Ralf Drautz}
\email{ralf.drautz@rub.de}
\affiliation{Interdisciplinary Centre for Advanced Materials Simulation, Ruhr-Universit{\"a}t Bochum, 44801 Bochum, Germany}

\date{\today}

\begin{abstract}
We devise automated workflows for the calculation of Helmholtz and Gibbs free energies and their temperature and pressure dependence and provide the corresponding computational tools. We employ non-equilibrium thermodynamics for evaluating the free energy of solid and liquid phases at a given temperature and reversible scaling for computing free energies over a wide range of temperatures, including the direct integration of $P$-$T$ coexistence lines. By changing the chemistry and the interatomic potential, alchemical and upscaling free energy calculations are possible. Several examples illustrate the accuracy and efficiency of our implementation.
\end{abstract}

\maketitle

\section{Introduction}

Free energies are crucial for thermodynamic analysis and provide valuable insight into the relative stability of phases and their coexistence.
The calculation of free energies from atomistic simulations is far from trivial, also because free energies cannot be expressed easily as thermodynamic averages that may be obtained in atomistic simulations directly. \\

Thermodynamic integration \cite{Kirkwood1935,Smit2001} is widely employed to compute free energies. 
In thermodynamic integration, the system of interest is related to a reference system with known free energy. 
A parameter switches the energy of the system of interest to the reference system continuously and smoothly.
For example, the free energy of solids can be computed by transforming from a non-interacting Einstein crystal to a system of interest \cite{Frenkel1984}.
For liquids, reference systems such as the ideal gas \cite{Hermans1988,Mei1992}, or pair potentials like Lennard-Jones \cite{Harvey2011} or Uhlenbeck-Ford \cite{PaulaLeite2016} are utilised. \\

Thermodynamic integration is computationally intensive because many separate calculations are required for discrete points along the non-physical path that connects the reference system to the system of interest \cite{Smit2001}. 
Non-equilibrium approaches to thermodynamic integration \cite{Watanabe1990}, in which the system of interest is transformed to the reference system as a function of time, lead to significant improvements in computational cost and efficiency.
Non-equilibrium calculations for the computation of Helmholtz free energies have been applied to solids \cite{Freitas2016} and liquids \cite{PaulaLeite2019} within the framework of the molecular dynamics code \textsc{lammps} \cite{Plimpton1995}.
Employing reversible scaling \cite{deKoning1999}, non-equilibrium calculations are carried out over thermodynamic state variables by relating the switching parameter to temperature. 
In this way, the variation of the Helmholtz free energy with temperature can be computed from a single non-equilibrium calculation starting from the Helmholtz energy at a reference temperature. 
By scaling temperature and pressure simultaneously, $P$-$T$ phase boundaries can be computed within a single simulation \cite{deKoning2001}.\\

In practical applications of non-equilibrium methods, several simulations need to be combined, and parameters need to be set, which makes the computational procedure cumbersome for non-specialists. Therefore, automated protocols that efficiently carry out non-equilibrium thermodynamic integration and establish a bridge from atomistic simulations to thermodynamics are highly desirable.

A general workflow can be subdivided into four broad steps:
\begin{enumerate}
\item evaluation of basic properties such as volume or pressure for the system of interest;
\item setting of reference system parameters to resemble the system of interest as closely as possible;
\item time-dependent switching between the system of interest and the reference system to compute the free energy; 
\item temperature sweep to calculate the temperature dependence of the Helmholtz or Gibbs free energy for constant volume, pressure or along pressure-temperature paths (i.e. for phase coexistence). 
\end{enumerate}

Several works implement parts of this workflow, but a general, automated approach is not available. This limits the widespread application of non-equilibrium methods for the computation of thermodynamic properties.\\

Here, we present an entirely automated workflow implementation, requires only minimal input and can be used to calculate both Helmholtz and Gibbs free energies. 
The workflow has four main applications: (i) calculation of the Helmholtz and Gibbs free energy at a given temperature, (ii) Helmholtz/Gibbs free energy calculation as a function of temperature at constant volume/pressure, (iii) calculation of the pressure-temperature coexistence line of two phases, and (iv) free energy computation for alchemical changes and upscaling. 
The workflow is suitable for single and multi-component systems, but configurational contributions to the free energy are not evaluated.\\

We demonstrate our automated workflow for calculating the pressure-temperature phase diagram of Ti using an embedded atom method (EAM) interatomic potential \cite{Mendelev2016}. 
In particular, we use three alternative methods that our workflow implements for the computation of the phase diagram. For multi-component applications, we demonstrate the Helmholtz free energy calculation of a binary CuZr system \cite{Mendelev2009}.  
Also, we present an algorithm for alchemical changes and upsampling in which the chemistry of the system or the interatomic potential in use is continuously transformed. 
We demonstrate upsampling by transforming between a relatively computationally inexpensive EAM potential for Cu \cite{Mishin2001}, and the more expensive, recently developed atomic cluster expansion (ACE) potential \cite{Lysogorskiy2021,Drautz2019}. 
Upsampling speeds up the free energy calculations by a factor of five, without loss of precision. 
Furthermore, in the CuZr system, we employ alchemical transformations to integrate from Cu to Zr.\\

The remainder of the paper is organised as follows: in Section \ref{sec:nehi}, we discuss the non-equilibrium calculation of free energy, followed by the temperature dependence of free energy in Section \ref{sec:temp-dependence}, and in Section \ref{sec:alchemy}, an algorithm for alchemical changes and upsampling is discussed. Finally, we demonstrate the application of the algorithms in Section \ref{sec:examples}, discuss the practical implementation of the algorithms in the form of a Python program in Section \ref{sec:implementation}, and conclude in Section \ref{sec:conclusion}.

\section{Non-equilibrium calculation of free energy differences} \label{sec:nehi}

We assume that the Helmholtz free energy for an initial Hamiltonian $H_i$ is known and the target is the computation of the free energy for a final Hamiltonian $H_f$. 
To this end, the two Hamiltonians are combined into the Hamiltonian $H(\lambda)$ with a parameter $\lambda$ that continuously switches between the initial and final Hamiltonian,
\begin{equation}
H_i = H (\lambda_i) \,\,\, \textrm{and} \,\,\, H_f = H (\lambda_f) \,.
\end{equation}
The reversible work for switching along $\lambda$ is given by \cite{Smit2001}
\begin{equation} \label{eq:2a}
  W^{rev}_{i \to f} = \int_{\lambda_i}^{\lambda_f} 
  \bigg \langle \frac{\partial H (\lambda)}{\partial \lambda} \bigg \rangle_{\lambda '} d\lambda' \,.
\end{equation}
If $\lambda$ is varied as a function of time, the work done is given by,
\begin{equation} \label{eq:2b}
  W^{s}_{i \to f} = \int_{t_i}^{t_f}
  \frac{d \lambda(t)}{dt}
  \frac{\partial H (\lambda)}{\partial \lambda}  dt \,,
\end{equation}
where $W^{s}$ is the dynamic work done along the process, $\lambda_i = 
\lambda(t_i)$ and $\lambda_f = \lambda(t_f)$. The time over which $\lambda$ is switched, $t_f-t_i$, is the switching time $t_{sw}$.
The free energy difference is related to the dynamic work as,
\begin{equation} \label{eq:df2}
  \Delta F = W^{rev} = {{W}^{s}} - {{E}^{d}} \,,
\end{equation}
with ${E}^{d}$ being the average dissipated energy. The energy dissipation depends on the rate at which the Hamiltonian is switched, with ${E^d} \to 0$ for $t_{sw} \to \infty$. If the transformation is slow and close to an ideal quasistatic process, the dissipated energy when switching the system from the initial to the final state is the same as for switching from the final to the initial state \cite{deKoning2005} 
\begin{equation}
E_d = ~{E}_{i \to f}^{d} = {E}_{f \to i}^{d}  \,,
\end{equation}
and therefore
\begin{equation}
  \Delta F = \frac{1}{2} [W_{i \to f}^{rev} - W_{f \to i}^{rev}]
  = \frac{1}{2} [{W}_{i \to f}^{s} - {W}_{f \to i}^{s}]    \,.\label{eq:dfl}
\end{equation}
The magnitude of energy dissipation follows as,
\begin{equation} \label{eq:ed}
  {E}^{d} = \frac{1}{2} [{W}_{i \to f}^{s} + {W}_{f \to i}^{s}] \,.
\end{equation}

For the computation of the Gibbs free energy difference in the isobaric ensemble only small modifications are necessary. We relate the initial and final Hamiltonians with parameter $\lambda$ as before and include the dependence of pressure on $\lambda$, $P(\lambda)$, with
\begin{equation}
P_i = P (\lambda_i) \,\,\, \textrm{and} \,\,\, P_f = P (\lambda_f) \,.
\end{equation}
The work done by switching is then obtained as
\begin{equation} \label{eq:2a_1}
  W^{s} = \int_{t_i}^{t_f}
  \frac{d \lambda}{dt}
  \left( \frac{\partial H (\lambda)}{\partial \lambda} +  \frac{\partial P (\lambda) V }{\partial \lambda}  \right)  dt \,,
\end{equation}
and for a quasistatic process
\begin{equation}
  \Delta G = \frac{1}{2} [{W}_{i \to f}^{s} - {W}_{f \to i}^{s}]  \,.\label{eq:dfl3}
\end{equation}

\section{Temperature dependence of the free energy}  \label{sec:temp-dependence}

The temperature dependence of the free energy is computed in two steps. First, the free energy difference between a reference system and the system of interest at constant temperature $T_i$ is obtained. 
In a second step, the free energy of the system of interest at $T_i$ is taken as the starting point for a temperature sweep for the computation of the free energy in the interval from $T_i$ to $T_f$. We implement temperature sweeps
\begin{enumerate}
    \item at constant volume,
    \item at constant pressure,
    \item along the pressure-temperature two-phase coexistence line.
\end{enumerate}

\subsection{Free energy at constant temperature}

The starting point is a reference Hamiltonian $H_i$ for which the free energy $F_i(N,V,T)$ is known. 
For the solid we use an Einstein crystal and for the liquid an ideal gas combined with the Uhlenbeck-Ford model \cite{PaulaLeite2016}. 
In the Einstein crystal, the atoms are bound to reference positions, which means that the free energy is computed for the given atomic configuration and the configurational entropy is not included. 
This is different for the liquid reference. In the liquid, atoms are free to move and exchange and the configurational entropy is part of the calculation. 
The free energy of the reference systems is summarized in Appendix~\ref{app:Hi}. 
The combined Hamiltonian is then written as
\begin{equation}
\label{eq:H_lambda}
H(\lambda(t)) = ( 1 - \lambda(t)) H_i + \lambda(t) H_f \,,
\end{equation}
and the integration over time is carried out from $\lambda(t_i) = 0$ to $\lambda(t_f) = 1$. 
The free energy of the system of interest is obtained by switching $\lambda$ at constant temperature and volume,
\begin{equation} \label{eq:deltaf}
F_f(N,V,T) = F_i(N,V,T) + \Delta F \,.
\end{equation}
where $\Delta F$ is computed using Eq.~\eqref{eq:dfl}.
As the pressure for $F_f(N,V,T)$ can be directly obtained, the Gibbs free energy may be calculated as $G_f = F_f + PV$.
The evaluation of the free energy at constant temperature from a known reference is implemented in algorithm~\ref{algo:step1}.

\begin{algorithm}[H]
    \caption{ \label{algo:step1} Compute free energy at constant $T$}
    \begin{algorithmic}[1]
    \State calculate $V$ at $(NPT)$ for $H_f$
    \If{solid}
     \For {all atoms}
        \State calculate: 
            \State average mean squared displacement $\av{ (\Delta \pmb{r})^2}$ 
            \State spring constant $k$
     \EndFor
     \State setup reference $H_i = H_E$ (see Appendix \,\ref{app:Hi:solid})
    \ElsIf{liquid}
      \State calculate density $\rho$
      \State setup reference $H_i = H_{UF}$ (see Appendix \,\ref{app:Hi:liquid})
    \EndIf
    \For {$n$ independent runs}
      \State equilibrate for time $t_{eq}$ 
      \State switch $ \lambda: 0 \rightarrow 1$ over time $t_{sw}$
      \State calculate work ${W}_{i \to f}^{s}$ (Eq.~(\ref{eq:2b}))
      \State equilibrate  for time $t_{eq}$ 
      \State switch $ \lambda: 1 \rightarrow 0$ over time $t_{sw}$
      \State calculate work ${W}_{f \to i}^{s}$ (Eq.~(\ref{eq:2b}))
     \EndFor
     \State average over $n$ independent runs $\Delta F =  \frac{1}{2} [{W}_{i \to f}^{s} - {W}_{f \to i}^{s}]$
     \State calculate free energy    
     \State$F_f(N,V,T) = F_i(N,V,T) + \Delta F$
     \If {$P$ is known}
        \State $G_f(N,V,T) = F_f(N,V,T) + PV$
    \EndIf
    \end{algorithmic} 
\end{algorithm}

\subsection{Temperature sweep}

Next the free energy obtained in the previous section is taken as the initial free energy $F_i$ at the temperature $T_i$ and volume $V_i$. 
We employ reversible scaling\cite{deKoning1999} to sweep the temperature at constant volume, constant pressure or along a $P$-$T$ phase boundary.

Apart from $F(N,V,T)$ and $G(N,P,T)$, by numerical differentiation entropy

\begin{equation} \label{eq:cp1}
S = -\left( \frac{dG}{dT} \right)_P
\end{equation}

and specific heat 

\begin{equation} \label{eq:cp2}
C_P = T\left( \frac{dS}{dT} \right)_P
\end{equation}

are obtained.


\subsubsection{Constant volume}

For sweeping the temperature at constant volume we use the relation
\begin{equation}
F(N,V,T_f) =  F(N,V,T_i) - \frac{3}{2} k_\mathrm{B} T_f N \ln   \frac{T_f}{T_i}+   \frac{T_f}{T_i} \Delta F  \,, \label{eq:dF} 
\end{equation}
where $\Delta F$ is obtained from scaling the Hamiltonian at constant temperature (Eq.~(\ref{eq:wa1})). The derivation of this expression is summarized in Appendix~\ref{app:revscale}. The temperature sweep is implemented in algorithm~\ref{algo:step2}. 

\begin{algorithm}[H]
    \caption{ \label{algo:step2} $T$ sweep for constant $V$ or $P$}
    \begin{algorithmic}[1]
        \If {constant $V$}
            \State $F(N,V,T_i)$ from algorithm \ref{algo:step1}
        \ElsIf {constant $P$}
            \State $F(N,V,T_i)$ from algorithm \ref{algo:step1}
            \State calculate $G(N,P,T_i) =  F(N,V,T_i) + PV_i$
        \EndIf  
        \For {$n$ independent runs}
          \If {constant $V$}
              \State equilibrate for time $t_{eq}$ in NVT ensemble
              \State switch $ \lambda: 1 \rightarrow T_i/T_f$ over time $t_{sw}$ 
              \State calculate work ${W}_{i \to f}^{s}$ (Eq.~(\ref{eq:wa1}))
              \State equilibrate  for time $t_{eq}$ in NVT ensemble
              \State switch $ \lambda: T_i/T_f \rightarrow 1$ over time $t_{sw}$
              \State calculate work ${W}_{f \to i}^{s}$ (Eq.~(\ref{eq:wa1}))
          \ElsIf {constant $P$}
              \State equilibrate for time $t_{eq}$ in NPT ensemble
              \State switch $ \lambda: 1 \rightarrow T_i/T_f$ over time $t_{sw}$ 
              \State calculate work ${W}_{i \to f}^{s}$ (Eq.~(\ref{eq:wa2}))
              \State equilibrate  for time $t_{eq}$ in NPT ensemble
              \State switch $ \lambda: T_i/T_f \rightarrow 1$ over time $t_{sw}$
              \State calculate work ${W}_{f \to i}^{s}$ (Eq. \ref{eq:wa2})
          \EndIf
        \EndFor
        \If {constant $V$}
           \State average over $n$ independent runs $\Delta F =  \frac{1}{2} [{W}_{i \to f}^{s} - {W}_{f \to i}^{s}]$ 
            \State calculate $F(N,V,T_f) =  F(N,V,T_i) - \frac{3}{2} k_\mathrm{B} T_f N \ln   \frac{T_f}{T_i}+   \frac{T_f}{T_i} \Delta F$
        \ElsIf {constant $P$}
            \State average over $n$ independent runs $\Delta G =  \frac{1}{2} [{W}_{i \to f}^{s} - {W}_{f \to i}^{s}]$ 
            \State calculate $G(N,P,T_f) =  G(N,P,T_i) - \frac{3}{2} k_\mathrm{B} T_f N \ln   \frac{T_f}{T_i}+   \frac{T_f}{T_i} \Delta G$
        \EndIf
        \State calculate $S$ and $C_P$ using Eq.~(\ref{eq:cp1}) and Eq.~(\ref{eq:cp2}).
    \end{algorithmic} 
\end{algorithm}

\subsubsection{Constant pressure}

For calculations at constant pressure, the Gibbs free energy reference is obtained as $G_i = F_i + P V_i$, where $P$ is the pressure at volume $V_i$. We then use
\begin{equation}
G(N,P,T_f) =  G(N,P,T_i) - \frac{3}{2} k_\mathrm{B} T_f N \ln   \frac{T_f}{T_i}+   \frac{T_f}{T_i} \Delta G  \,, \label{eq:dG} 
\end{equation}
where $\Delta G$ is obtained from scaling Hamiltonian and pressure at constant temperature (Eq.~(\ref{eq:wa2})), see Appendix~\ref{app:revscale}.
The temperature sweep is implemented in algorithm~\ref{algo:step2}.

\subsubsection{$P$-$T$ coexistence line} \label{sec:dcc}

For sweeping temperature $T$ along the coexistence line $P(T)$ first an initial coexistence point between two phases $\alpha$ and $\beta$ is established, $G_{\alpha}(N, P_i, T_i) = G_{\beta}(N, P_i, T_i)$. 
Then scaling temperature and adapting pressure to continuously fulfill the Clausius-Clapeyron condition, a series of coexistence points is obtained.
The necessary equations are summarized in Appendix~\ref{app:revscale}. The workflow is detailed in algorithm~\ref{algo:dcc}.

\begin{algorithm}[H]
    \caption{ \label{algo:dcc} $T$ sweep along $P(T)$ coexistence line}
    \begin{algorithmic}[1]
    \For {system in $\alpha$, $\beta$}
      \State for $P_i > 0$ calculate $G(N,P,T_i)$ for initial temperature $T_i$ from algorithm \ref{algo:step1}
      \State calculate $G(N, P_i, T)$ for various temperatures from algorithm \ref{algo:step2}
    \EndFor
    \State for $P_i$ determine $T_i$ such that $G_{\alpha}(N, P_i, T_i) = G_{\beta}(N, P_i, T_i)$
    \For {system in $\alpha$, $\beta$}
      \For {$n$ independent runs}
          \State equilibrate for time $t_{eq}$ in $NPT$ ensemble
          \State switch $ \lambda: 1 \rightarrow T_i/T_f$ over time $t_{sw}$
          \State  calculate pressure ${\Delta P}_{i \to f}^{s}$
          \State  calculate work ${W}_{i \to f}^{s}$
          \State equilibrate  for time $t_{eq}$ in $NPT$ ensemble
          \State switch $ \lambda: T_i/T_f \rightarrow 1$ over time $t_{sw}$
          \State  calculate pressure ${\Delta P}_{f \to i}^{s}$
          \State  calculate work ${W}_{f \to i}^{s}$
      \EndFor
    \EndFor
    \State average $n$ independent runs $\Delta P = \frac{1}{2} [{\Delta P}_{i \to f}^{s} - {\Delta P}_{f \to i}^{s}]$
    \State average $n$ independent runs $W = \frac{1}{2} [{W}_{i \to f}^{s} - {W}_{f \to i}^{s}]$
    \State calculate $P_f = (T_f/T_i) (P_i + \Delta P$)
    \State calculate $G(N,P_f,T_f) =  G(N,P_i,T_i) - \frac{3}{2} k_\mathrm{B} T_f N \ln   \frac{T_f}{T_i} +   \frac{T_f}{T_i} \Delta G$
    \State with $ G_{\alpha}(N,P_f,T_f) = G_{\beta}(N,P_f,T_f) = G(N,P_f,T_f)$
    \end{algorithmic} 
\end{algorithm}

\section{Alchemical changes and upsampling}  \label{sec:alchemy}

For efficient defect formation free energies or for the computation of phase diagrams, alchemical changes and upsampling are useful.
To this end, we provide algorithm~\ref{algo:alchemical} which continuously transforms atoms and atomic interactions from an initial system to the final system. 
Along the transformation path, each atom may change its chemistry as described by the potential energy and mass. 
The integration is separated into two steps. 
First, we evaluate the free energy difference between the initial and final system by changing atomic interactions along $\lambda$, but at constant atomic masses. We transform according to Eq.~(\ref{eq:H_lambda}) and 
the integration over time is carried out from $\lambda(t_i) = 0$ to $\lambda(t_f) = 1$. 
The free energy difference, $\Delta F$, is then obtained from Eq. \eqref{eq:dfl}. 
In the second step, we consider the free energy change originating from the change in atomic masses.
This contribution is given by,

\begin{equation}
    \Delta F_{\mathrm{mass}} = \frac{3}{2} k_\mathrm{B} T \sum_{i=1}^N \ln{\bigg(\frac{m_i^{(i)}}{m_i^{(f)}}\bigg)} \,,
\end{equation}
as briefly summarized in Appendix~\ref{app:mass}.

Given the free energy $F_i$ of the initial system, the free energy of the final system is given by,
\begin{equation}
    F_f = F_i + \Delta F + \Delta F_{\mathrm{mass}}\,.
\end{equation}

If only the interatomic potential is changed along the path and the atomic masses remain constant, only the first step is necessary and the algorithm may be used for efficiently computing the free energy by starting from a less refined model of the same chemistry, similar in spirit to the upsampling in the TU-TILD approach \cite{Duff2015}.
 
\begin{algorithm}[H]
    \caption{ \label{algo:alchemical} Alchemical changes}
    \begin{algorithmic}[1]
    \State define initial and final chemistry for each pair of atoms
    \State define initial and final potential
    \State set up  $H_i$ and $H_f$   
    \For {$n$ independent runs}
      \State equilibrate for time $t_{eq}$
      \State switch $ \lambda: 0 \rightarrow 1$ over time $t_{sw}$
      \State  calculate work ${W}_{i \to f}^{s}$
      \State equilibrate  for time $t_{eq}$
      \State switch $ \lambda: 1 \rightarrow 0$ over time $t_{sw}$
      \State calculate work ${W}_{f \to i}^{s}$
   \EndFor
   \If {constant $V$}
     \State average $n$ independent runs $\Delta F =  \frac{1}{2} [{W}_{i \to f}^{s} - {W}_{f \to i}^{s}]$ 
     \State calculate    $F_f(N,V,T) = F_i(N,V,T) + \Delta F + \Delta F_\mathrm{mass}$
    \ElsIf {constant $P$}
     \State average $n$ independent runs $\Delta G =  \frac{1}{2} [{W}_{i \to f}^{s} - {W}_{f \to i}^{s}]$ 
     \State calculate $G_f(N,P,T) = G_i(N,P,T) + \Delta G + \Delta G_\mathrm{mass}$
   \EndIf 
    \end{algorithmic} 
\end{algorithm}

\section{Applications}  \label{sec:examples}

\subsection{Convergence with system size and switching time}

\begin{figure*}[ht!]
\centering
\includegraphics[width=0.6\textwidth]{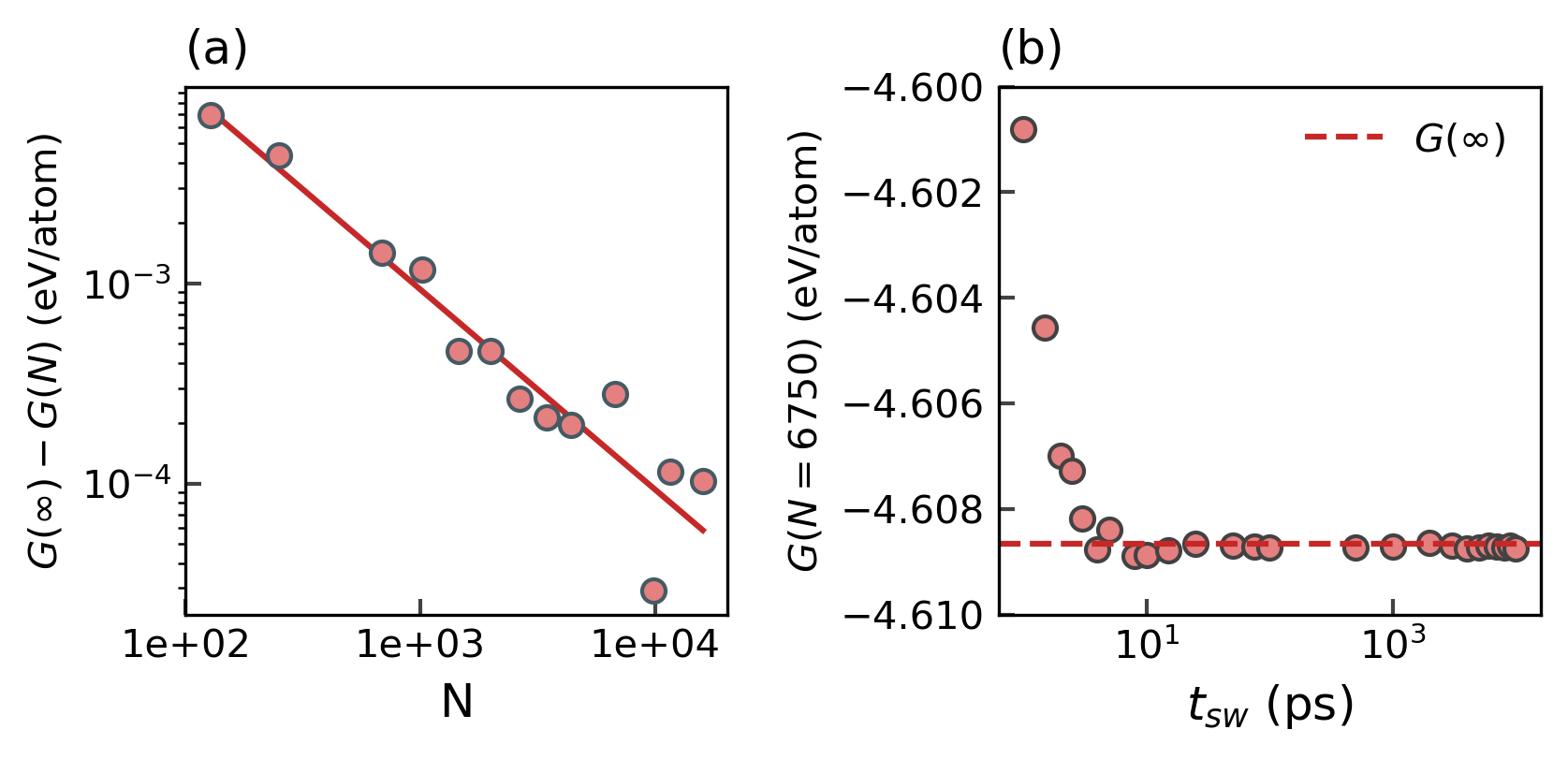}
\caption[Ti phase diagram]{\label{fig:conv} 
(a)  $G(\infty)-G(N)$ at 1000 K and zero pressure as a function of system size from 128 to 16000 atoms and a switching time of 100 ps. (b) $G(N=6750)$ calculated for each switching time. $G(\infty)$ is shown in dashed red line. Switching times from 1 ps to 10 ns are used for 6750 atoms.
}
\end{figure*}

For analysing the dependence of the free energy on system size and switching time we use bcc Fe with an EAM potential \cite{Meyer1998}. 
We choose this particular system to facilitate comparison with \citet{Freitas2016}. 
We calculate the free energy at 1000~K and zero pressure for the bcc structure using algorithm \ref{algo:step1} for various system sizes ranging from 128 to 16000 atoms. 
The switching is carried out over 100 ps. 
The calculated free energy, $G(N)$,  converges as a function of $1/N$, where $N$ is the number of atoms in the system \cite{Polson2000}.
We can thus evaluate $G(\infty)$, the free energy at the thermodynamic limit through an asymptotic analysis of $G(1/N)$.
The difference in free energy, $G(\infty)-G(N)$, is shown in Fig. \ref{fig:conv} (a), after taking into account corrections due to the fixed centre of mass \cite{Polson2000}.
From the figure, 3000 atoms are sufficient to obtain the free energy within an accuracy of 0.01 meV/atom. 

In Fig. \ref{fig:conv} (b), we show the free energy with varying switching time $t_{sw}$ from 1 ps to 10 ns. We use a bcc cell with $N=6750$ atoms at a temperature of 1000~K and zero pressure. Even for the relatively short switching time of 50 ps, the free energy can be evaluated with a precision of 0.1 meV/atom. At 100 K, we obtain $G(P=0, T=100 \mathrm{K}) = -4.263118(4)$ eV/atom over  $t_{sw}=10$ ns, which is in excellent agreement with \citet{Freitas2016}.

\subsection{Pressure-temperature phase diagram for Ti}

To illustrate the robustness and efficiency of the algorithms, we calculate the pressure-temperature phase diagram of Ti using an EAM potential~\cite{Mendelev2016}. 
We consider the bcc, hcp and liquid phases in the pressure range of 0-5 GPa and temperature range of 500-3000 K. The simulation cell consists of 4394 atoms for the bcc structure and 8878 atoms for the hcp and liquid phases. 

\begin{figure}[ht!]
\centering
\includegraphics[width=0.40\textwidth]{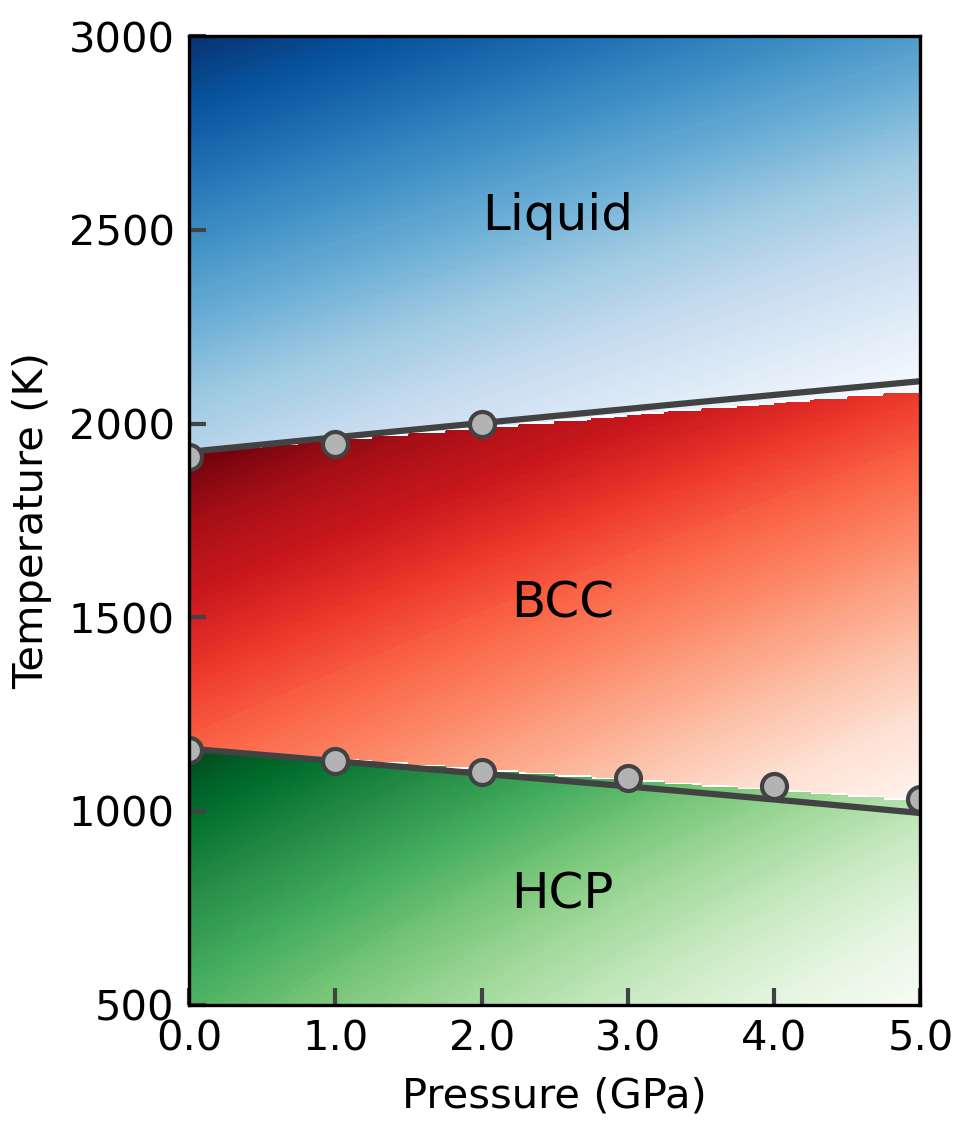}
\caption[Ti phase diagram]{\label{fig:tpdiagram} 
The pressure-temperature phase diagram for Ti using an EAM potential~\cite{Mendelev2016}. The regions of stability for bcc, hcp, and liquid are shown in red, green, and blue, respectively. The gradients in the color indicate decreasing free energy; darker color indicates decreasing free energy. Coexistence lines calculated using algorithm \ref{algo:dcc} are shown in grey. The coexistence line is further verified by using algorithm \ref{algo:step1}, the results of which are shown as grey circles. The three independent computations are in excellent agreement.
}
\end{figure}

Using the algorithms introduced in this work, we calculate the phase diagram using three different strategies. 
(i) We localize points of pairwise identical Gibbs free energy by sweeping temperature at constant pressure. This includes three steps:

\begin{itemize}
    \item Using algorithm \ref{algo:step1} for the hcp and bcc phase, we calculate $G(P,T)$ at $T=500$ K and pressures from 0 to 5 GPa in intervals of 0.25 GPa. For the liquid phase, $G(P,T)$ is calculated over the same pressure range, but at a temperature of 1500 K. For all calculations, a switching time of 50 ps was used.
    \item Starting from $G(P,T)$ calculated in the previous step, we follow algorithm \ref{algo:step2} to compute the dependence of the free energy on the temperature. For bcc and liquid phases, we scale the Hamiltonian of the system such that a temperature range up to 3000 K is covered. For hcp, the temperature range until 1500 K is traversed as the structure is unstable at higher temperatures.
    \item At each pressure, from the crossings of the free energies as a function of temperature, the phase transition temperature is located. 
\end{itemize}
The calculated phase diagram is shown in Fig. \ref{fig:tpdiagram}. The thermodynamic regions at which the bcc, hcp, and liquid phases are the most favourable energetically are red, green, and blue. 

(ii) We use algorithm \ref{algo:dcc} to sweep the $P$-$T$ coexistence lines directly. 
A prerequisite for algorithm \ref{algo:dcc} is a known coexistence point $(P_i, T_i)$ with $P_i > 0$. 
To this end, we use algorithm \ref{algo:step2} to calculate initial coexistence points $(P_i, T_i)$ at a low pressure of 0.01 GPa for bcc-hcp ($T=1158$~K) and bcc-liquid ($T=1931$~K). 
From these points, algorithm \ref{algo:dcc} is used to scale the temperature up to 2200~K for bcc-liquid coexistence and 1000~K for the bcc-hcp coexistence over a time of 1 ns. 
The calculated coexistence lines are shown in grey in Fig. \ref{fig:tpdiagram}. 
The coexistence lines using algorithm \ref{algo:dcc} show excellent agreement with coexistence points computed in (i).

(iii) We verify that the free energies along the coexistence lines are indeed identical by carrying out calculations using algorithm \ref{algo:step1}. The calculated coexistence points are shown in grey circles for both the bcc-hcp and the bcc-liquid coexistence lines. 

We predict the melting point of the bcc structure at zero pressure at 1913~K, in excellent agreement with the reported value (1918~K \cite{Mendelev2016}. 
We compute the hcp-bcc phase transition temperature to be 1150~K. 
Our calculations are in good agreement with direct molecular dynamics simulations (1150~K), and lattice switch Monte Carlo method (1152~K) \cite{Mendelev2016}. 
The $\omega$ phase, which appears in the experimental phase diagram \cite{Gu1994} ($P>2$~GPa, $T=0$~K), always has a higher free energy than the hcp structure as predicted by the EAM potential. 
In order to arrive at the complete phase diagram, only minimal user input, such as the required phases, and the corresponding temperature and pressure ranges are required.

\subsection{Phase diagram of Si}

The selection of an interatomic potential for a particular application requires the validation of the physical properties predicted by the model. 
For Si, a wide range of interatomic potentials were compared in order to assess their quality and transferability in Ref. \onlinecite{Lysogorskiy2019}. 
In addition, the calculation of phase diagrams can provide further insight into the quality of an interatomic potential.  
We calculate the pressure-temperature phase diagram of Si using different interatomic potentials.
We consider five different potentials: Stillinger-Weber (SW)  \cite{Stillinger1984}, angular dependent potential (ADP) \cite{Starikov2020}, spectral neighbour analysis potential (SNAP) \cite{Zuo2020}, Tersoff \cite{Tersoff1988}, and modified embedded atom method (MEAM) \cite{Lee2007}.
As an initial step to ascertain the validity of the potential, we calculate the melting temperature at zero pressure. We use 4096 atoms for both solid and liquid simulation cells and use a switching time of 50 ps.
For the Tersoff potential, the calculated melting temperature is very low ($< 1400$~K), while for the MEAM potential it is very high ($> 2000$~K), compared to the experimental value of 1687 K. Therefore we do not consider Tersoff and MEAM for further calculations.
The melting temperature for the SW potential is 1678~K, for ADP 1850~K and for SNAP 1405~K. 

\begin{figure}[t!]
\centering
\includegraphics[width=0.40\textwidth]{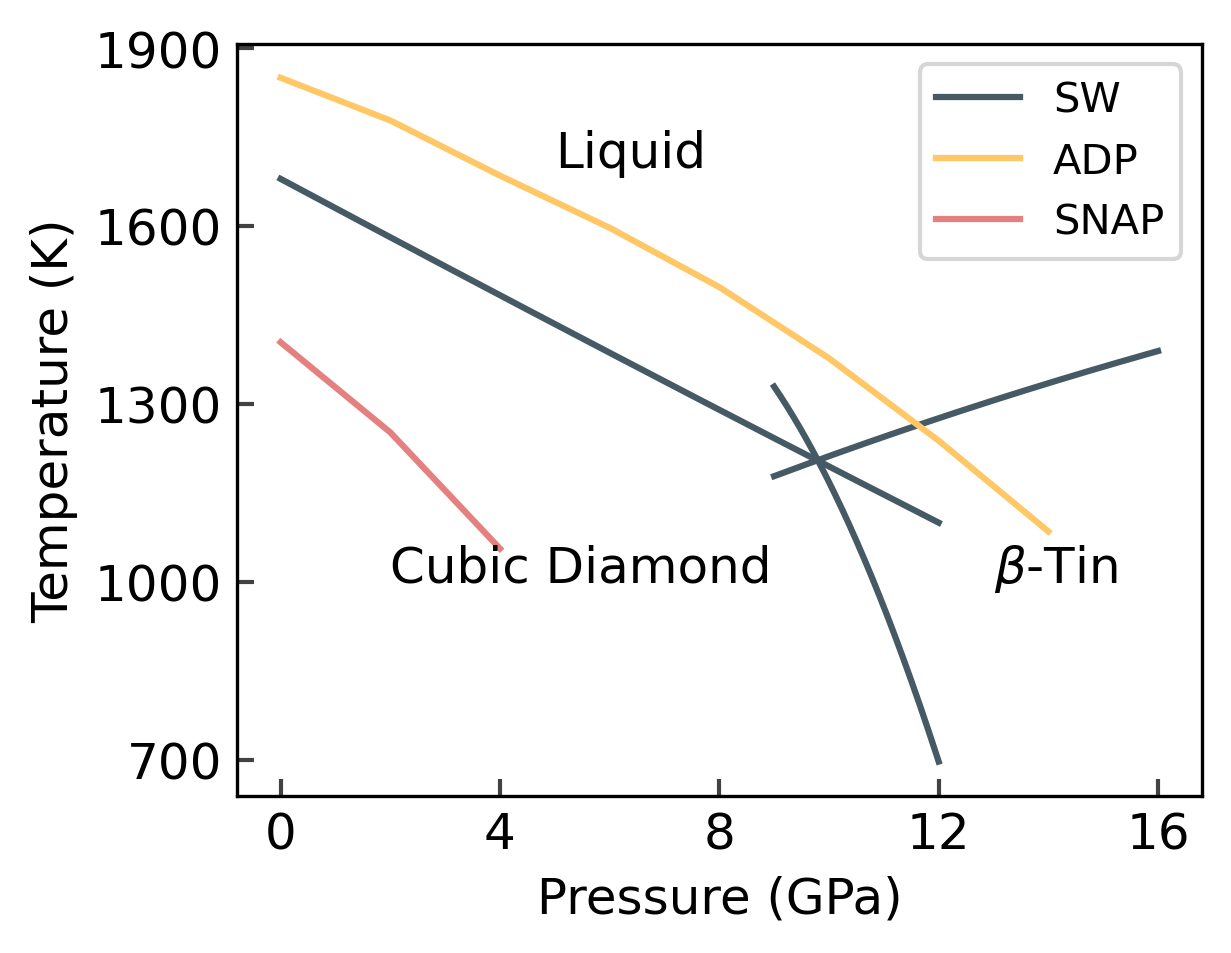}
\caption[Si phase diagram]{\label{fig:si} 
Pressure-temperature phase diagram of Si calculated using three different potentials. SW potential \cite{Stillinger1984}, ADP \cite{Starikov2020}, and SNAP \cite{Zuo2020} potentials are shown.}
\end{figure}

After calculating the melting temperature, we find the coexistence line between cubic diamond and liquid at various pressures. 
Additionally, we consider the $\beta$-tin phase, which is a high-pressure polymorph in Si, and we estimate the coexistence line for cubic diamond -- $\beta$-tin and $\beta$-tin -- liquid. 
The calculated phase diagram is shown in Fig. \ref{fig:si}. Note that we did not include the \textit{sc-16} phase which is known to be stabilized in the SW phase diagram \cite{Romano2014} in contrast to experiments.

As shown in Fig. \ref{fig:si}, the phase diagram predicted by the SW potential is in good agreement with previous results reported for the same potential \cite{Romano2014}. Furthermore, our calculations predict the liquid -- cubic diamond -- $\beta$-tin triple point at 9.8 GPa and 1205 K, in excellent agreement with previous calculations using the SW potential \cite{Romano2014}.

The ADP and SNAP potentials cannot predict coexistence of three phases, likely due to the exclusion of high-pressure phases during the development of the interatomic potentials. 
For ADP the only stable phases are cubic diamond and liquid and the system does not transform to $\beta$-tin at high pressure. 
The cubic diamond-liquid coexistence line is overestimated by about 200 K at the range of pressures considered. 
The SNAP also does not exhibit a phase transformation into the $\beta$-tin phase. 
The cubic diamond-liquid coexistence line is underestimated by about 300 K.
Thus, our algorithms can be employed for efficient calculation of phase diagrams as predicted by different interatomic potentials, providing valuable information about the transferability of the potential at various thermodynamic conditions.

\subsection{Calculation of specific heat}

Algorithm \ref{algo:step2} provides free energies as a function of temperature. 
Eq.~(\ref{eq:cp1}) and Eq.~(\ref{eq:cp2}) provide an efficient method to compute the specific heat. 
It is worthwhile to stress that there are no additional calculations required for $C_P$; it is directly available from algorithm~\ref{algo:step2}.

Alternatively, $C_P$ can also be calculated from the fluctuations in the isothermal-isobaric ensemble by

\begin{equation} \label{eq:cp3}
    \langle (\delta(U + PV))^2 \rangle_{NPT} = k_\mathrm{B}T^2 C_P \,,
\end{equation}

\noindent where $U$ is the internal energy. 

To illustrate the calculation of $C_P$, we use an EAM potential~\cite{Mishin2001} for Cu and compute the free energy as a function of the temperature using algorithm~\ref{algo:step2}. 
We use a simulation cell with 4000 atoms at zero pressure and a temperature range of 600-1200\,K. 
The switching is carried out over 1\,ns. 
To compare with  Eq.~(\ref{eq:cp3}), we further run MD simulations in the NPT ensemble with a system size of 4000 atoms over the same temperature range at intervals of 100 K. 
For each temperature, the MD simulation was run for 1\,ns, and five independent calculations were run for each temperature to estimate the error.

\begin{figure}[t!]
\centering
\includegraphics[width=0.40\textwidth]{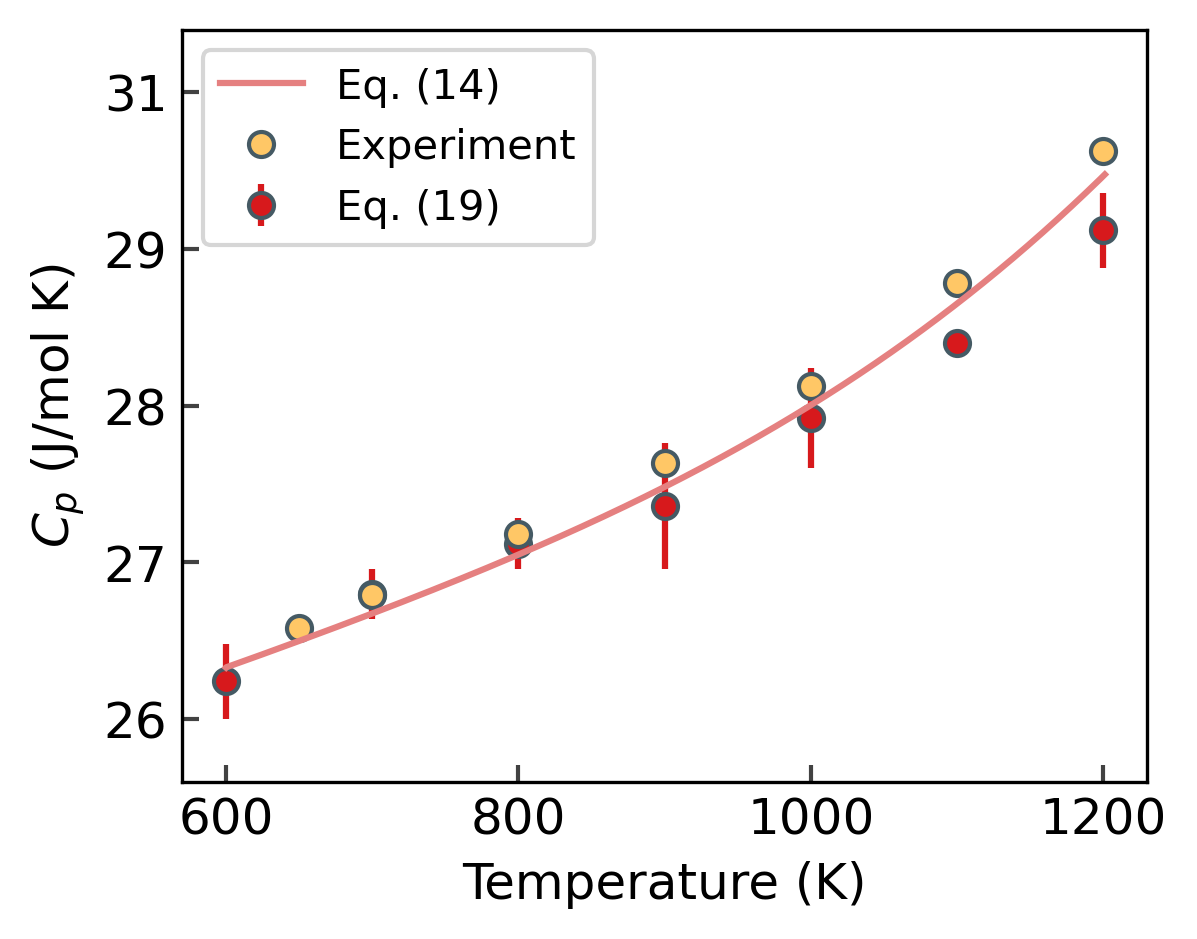}
\caption[cp]{\label{fig:cp} 
$C_P$ of Cu calculated using an EAM potential \cite{Mishin2001}. Orange line: algorithm \ref{algo:step2} and Eq.~(\ref{eq:cp2}), yellow circles: experiment~\cite{White1984}, red circles: MD calculations using Eq.~(\ref{eq:cp3}).}
\end{figure}

The results are shown in Fig.~\ref{fig:cp}. It is clear that both methods show fair agreement with each other and experiment. However, $C_p$ using Eq.~(\ref{eq:cp2}) is obtained from a single calculation compared to longer MD runs for each temperature using Eq.~(\ref{eq:cp3}). 

\subsection{Gibbs free energy of the CuZr system} \label{sec:cuzr}

\begin{figure}[t!]
\centering
\includegraphics[width=0.40\textwidth]{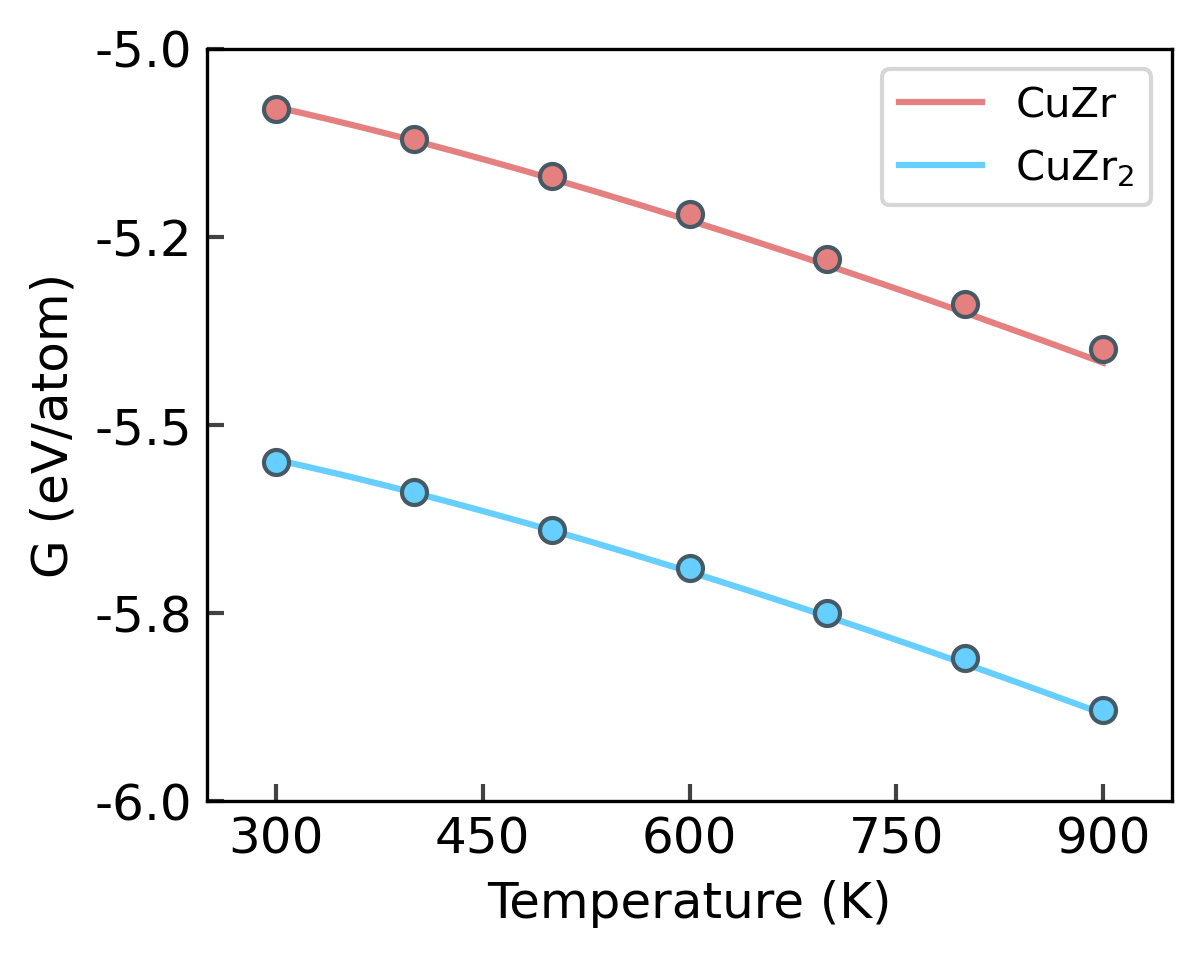}
\caption[Ti phase diagram]{\label{fig:cuzr} 
Free energy of the CuZr (red) and $\mathrm{CuZr}_2$ (blue). Solid line: free energy calculated using algorithm \ref{algo:step2}, circles: reference from \citet{Tang2012}. 
}
\end{figure}

The algorithms presented in this work are applicable to multi-component materials, but exclude the computation of configurational entropy in the solid phase. 
We use CuZr as an example and calculate the free energy for both CuZr (B2 structure) and $\mathrm{CuZr}_2$ ($\mathrm{C11}_b$ structure) using an EAM potential \cite{Mendelev2009}. 
The simulation cells for CuZr and $\mathrm{CuZr}_2$ contain 2000 and 1472 atoms, respectively. 
We calculate the free energy in the temperature range from 300-900 K and zero pressure for both structures using algorithm \ref{algo:step2}. 
The calculated free energy is shown in Fig.~\ref{fig:cuzr} and compared to the results from \citet{Tang2012}. 
Our calculations agree very well with the reported values. 
By employing algorithm~\ref{algo:step2}, however, it is possible to obtain the free energy over the whole temperature in a single simulation.

\subsection{Alchemical changes and upsampling}

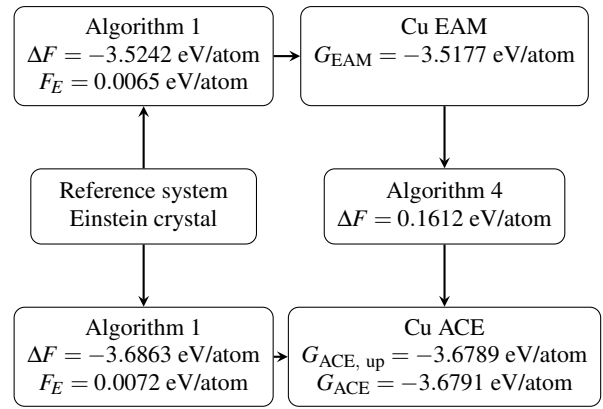
\begin{figure}[ht!]
\begin{tikzpicture}[node distance=2cm]
\node (a1) [startstop] {
\begin{tabular}{c}
Algorithm \ref{algo:step1}\\
$\Delta F=-3.5242$ eV/atom\\
$F_E=0.0065$ eV/atom\\
\end{tabular}
};
\node (a2) [startstop, below of=a1,] {
\begin{tabular}{c}
\small{Reference system}\\
Einstein crystal
\end{tabular}
};
\node (a3) [startstop, below of=a2] {
\begin{tabular}{c}
Algorithm \ref{algo:step1}\\
$\Delta F=-3.6863$ eV/atom\\
$F_E=0.0072$ eV/atom\\
\end{tabular}
};
\node (b1) [startstop, right of=a1, xshift=2cm] {
\begin{tabular}{c}
Cu EAM\\
$G_\mathrm{EAM}=-3.5177$ eV/atom\\
 \\
\end{tabular}
};
\node (b2) [startstop, right of=a2, xshift=2cm] {
\begin{tabular}{c}
Algorithm \ref{algo:alchemical}\\
$\Delta F=0.1612$ eV/atom\\
\end{tabular}
};
\node (b3) [startstop, right of=a3, xshift=2cm] {
\begin{tabular}{c}
Cu ACE\\
$G_{\mathrm{ACE,~up}}=-3.6789$ eV/atom\\
$G_\mathrm{ACE}=-3.6791$ eV/atom\\
\end{tabular}
};
\draw [arrow] (a2) -- (a1);
\draw [arrow] (a2) -- (a3);
\draw [arrow] (a1) -- (b1);
\draw [arrow] (a3) -- (b3);
\draw [arrow] (b1) -- (b2);
\draw [arrow] (b2) -- (b3);
\end{tikzpicture}
\caption{Illustration of the two routes by which the free energy for Cu within ACE can be calculated. Starting from the reference system, it can be directly calculated using algorithm \ref{algo:step1}. Alternatively, first the free energy of the EAM potential is calculated using algorithm \ref{algo:step1}, after which the system is upsampled to ACE using algorithm \ref{algo:alchemical}.} \label{fig:alchemy-layout}
\end{figure}

We demonstrate algorithm \ref{algo:alchemical} using two examples. In the first example, we use upsampling in which a system is transformed between two interatomic potentials. 
Such a scenario is akin to algorithm \ref{algo:step1}, albeit with a more complex reference system, and is similar to the TU-TILD approach \cite{Duff2015}. 
Here, we switch between a computationally cheap EAM potential and a relatively more expensive ACE potential. 
To obtain the free energy of Cu using the ACE potential at a temperature of 100 K and zero pressure, we follow two different routes as illustrated in Fig.~\ref{fig:alchemy-layout}. 
One approach is to evaluate the free energy directly starting from the reference Einstein crystal using algorithm~\ref{algo:step1}. 
In the second approach, we compute the free energy of the Cu EAM potential before the ACE potential. 
Compared to ACE, the EAM potential is about two orders of magnitude faster.
From the EAM potential, we use algorithm~\ref{algo:alchemical} to transform the system to the ACE potential. 
From the free energy difference, $\Delta F$, during this transformation, we can calculate the free energy of Cu for the ACE potential, by $G_\mathrm{ACE,~up} = G_\mathrm{EAM} + \Delta F$. 
As shown in Fig. \ref{fig:alchemy-layout}, we find that both routes arrive at the same result.

\begin{figure}[t!]
\centering
\includegraphics[width=0.40\textwidth]{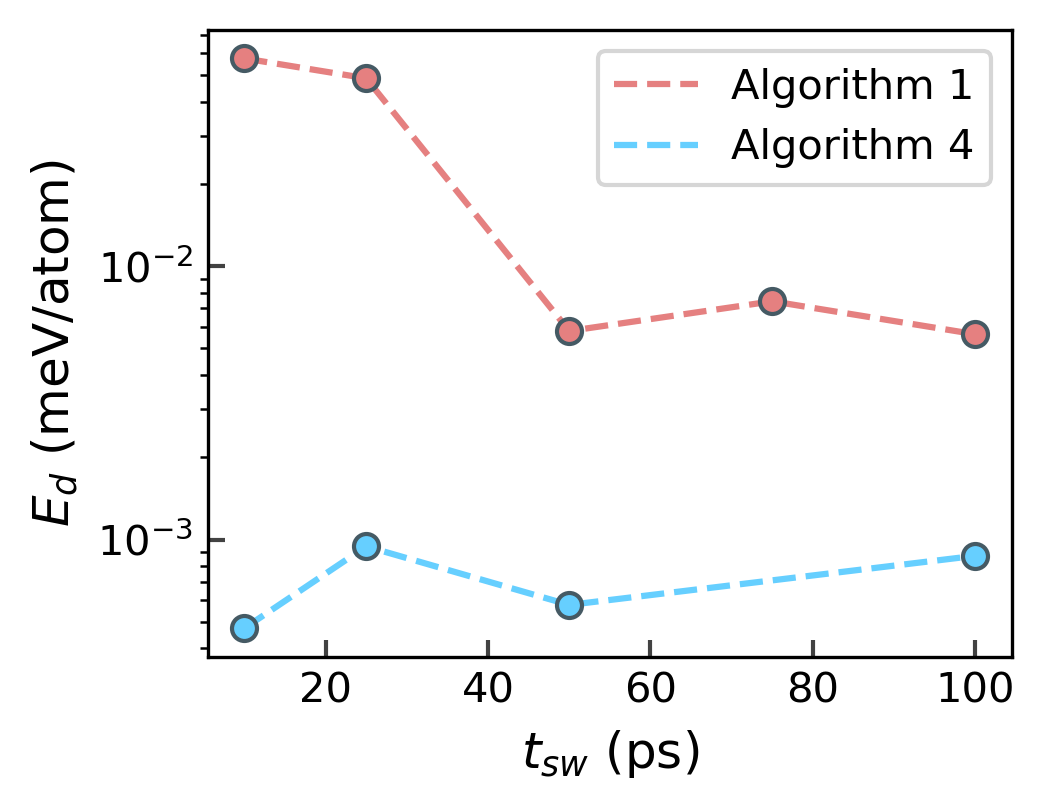}
\caption[Alchemical changes]{\label{fig:alchemy} 
Energy dissipation during the calculation of $G_\mathrm{ACE}$ using algorithm \ref{algo:step1} (red circles), and for switching between the EAM and ACE potentials (blue circles). 
}
\end{figure}

The advantage of using upsampling to estimate the free energy can be understood from Fig. \ref{fig:alchemy}. 
In Fig. \ref{fig:alchemy}, we show the energy dissipation, $E_d$ in work done during switching in the two routes discussed above: the energy dissipation during the calculation of $G_\mathrm{ACE}$ for switching time of 10--100\,ps is shown in red, the energy dissipation during upsampling is shown in blue. 
For all switching times considered, $E_d$ is lower for upsampling by at least an order of magnitude. 
The energy dissipation depends on the similarity of the two systems, and the Cu EAM potential is a more favourable reference state than the Einstein crystal. 
Thus, by using algorithm \ref{algo:step1} to compute the free energy for the computationally inexpensive potential, and by switching it to the more expensive one, it is possible to obtain free energies from switching times as low as 10\,ps. 
For a comparable accuracy, one needs at least 50\,ps of switching time for the direct calculation with the more expensive potential.

In the second example, we demonstrate the use of algorithm~\ref{algo:alchemical} for switching of chemical species. 
For the binary CuZr alloy in Section \ref{sec:cuzr}, we start in B2 structure at 800 K, which is the experimentally observed stable crystal structure at this temperature. 
We randomly swap 2\% of Zr atoms to Cu and use it as the initial state. 
We use algorithm \ref{algo:alchemical} to switch to 52 at. \% Zr.
This particular concentration range was chosen to compare the free energy of the system to that reported by \citet{Tang2012} using the same interatomic potential.
Along the integration path, the mass of the system is also transformed by adding a free energy contribution due to the change in kinetic energy as described in Appendix \ref{app:mass}. The system consists of 16000 atoms, and the alchemical switching is carried out over 100 ps. Additionally,  we evaluate the free energy at different concentrations at the same temperature using algorithm \ref{algo:step1}.

\begin{figure}[ht!]
\centering
\includegraphics[width=0.35\textwidth]{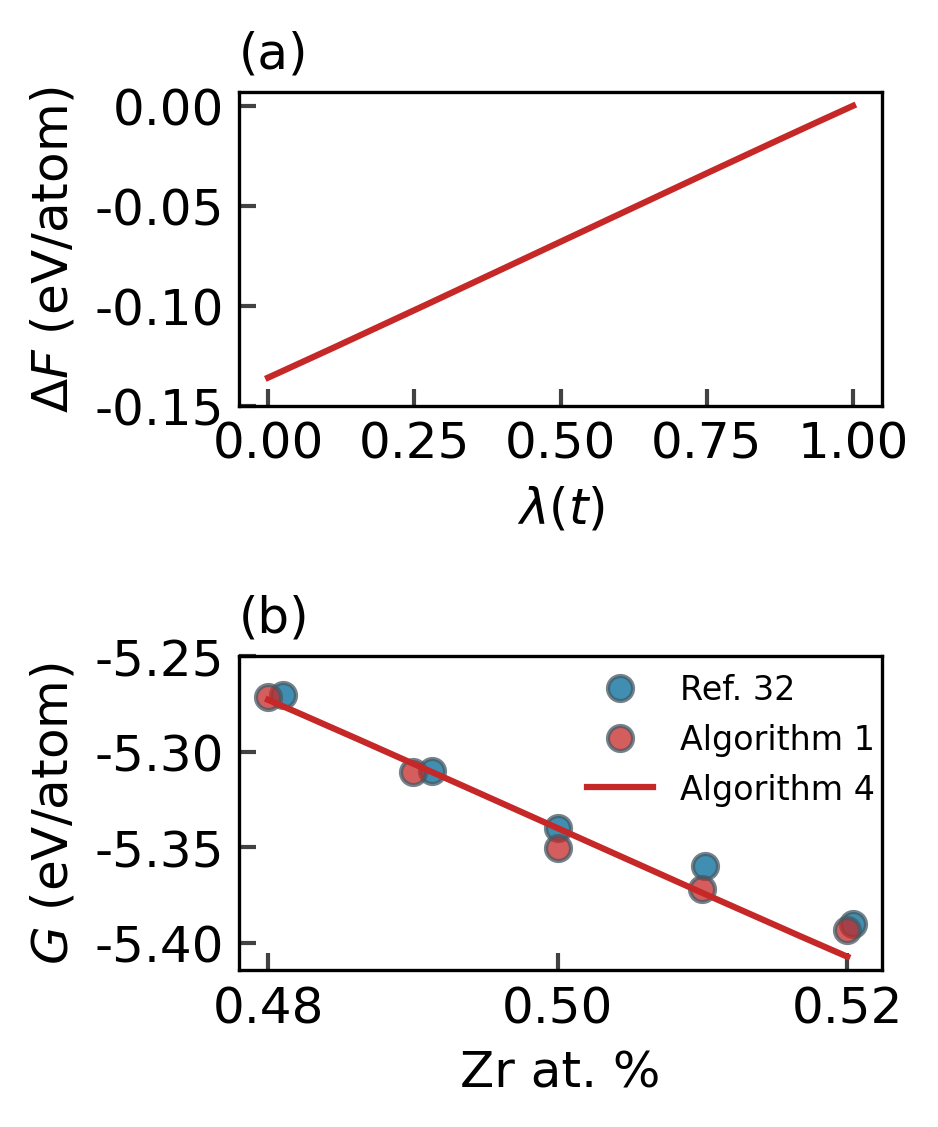}
\caption[Alchemy species]{\label{fig:alc-species} 
(a) $\Delta F$ as a function of $\lambda(t)$ along the alchemical integration path that switches a 48 at. \% Zr B2 structure to 52 at. \% Zr in the same lattice. (b) The free energy as a function of the Zr concentration calculated using algorithm \ref{algo:alchemical} (red line). Values reported by \citet{Tang2012} are shown as blue circles, while those calculated using Algorithm \ref{algo:step1} are shown as red circles.
}
\end{figure}

In Fig. \ref{fig:alc-species} (a), the cumulative reversible work along the integration path is shown, while in \ref{fig:alc-species} (b), the free energy as a function of concentration along the integration path is shown. 
We find good agreement with the free energy reported by \citet{Tang2012} and calculations using Algorithm \ref{algo:step1} even at intermediate points along the path (Zr at. \% = 48 to 52). 

\section{Implementation} \label{sec:implementation}

We provide a Python library, \textsc{calphy}, that includes all algorithms in this work. 
The library uses \textsc{lammps}~\cite{Plimpton1995} through the \textsc{pylammpsmpi} interface~\cite{Janssen2019, Janssen2021} to carry out the molecular dynamics simulations. 
The library also provides a command-line interface, where the necessary input options are provided through a text file. 
The input file used for the calculation of the free energy of CuZr system in the temperature range of 300-900~K at 0 pressure (Section \ref{sec:cuzr}) is shown in Fig. \ref{fig:inputfile}.

\begin{figure}
\centering
\begin{verbatim}
#elements
element: ['Zr', 'Cu']
#atomic mass
mass: [91.224, 63.546]
calculations:
#calculation mode: temperature-sweep
- mode: ts 
  #required temperature range
  temperature: [300, 900]
  #required pressure
  pressure: [0]
  #file containing input crystal structure
  lattice: [ZrCu.data]
  #state of the system
  state: [solid]
  #number of independent simulations 
  nsims: 3

md:
  #details of the inter-atomic potential
  pair_style: eam/fs
  pair_coeff: "* * ZrCu.eam.fs Zr Cu"
  #timestep for MD simulations
  timestep: 0.001
  #thermostat and barostat damping
  tdamp: 0.1
  pdamp: 0.1
  #timesteps for equilibration run
  te: 25000
  #timesteps for switching run
  ts: 50000
\end{verbatim}
\caption{ Input file for calculating $F(T)$ for the example discussed in Section \ref{sec:cuzr}.}
\label{fig:inputfile}
\end{figure}

The input file contains basic information about the system, such as the elements used in the calculations and atomic weights, followed by the required calculations and molecular dynamics options in separate blocks. 
In this case, we use algorithm \ref{algo:step2} to calculate the free energy over the given temperature range. 
The algorithm is specified under the  \verb|mode| keyword in the input file in the \verb|calculations| block. 
Further information, such as the temperature and pressure, is also provided in the same block. 
The input structure to be used for the calculation is read in the \textsc{lammps} data format. 
The input parameters pertaining to the molecular dynamics calculations are switched in the \verb|md| block, which contains information about the interatomic potential, timestep, thermostat, and barostat damping coefficients, as well as the switching and equilibration time. 
A detailed discussion of the input file is provided in the documentation of the library \cite{Documentation2021}.

The switching function $\lambda (t)$, which couples the system of interest and the reference system can have multiple functional forms as discussed in Appendix \ref{app:switching}.
In the case of solid systems, the switching between the system of interest and reference system is implemented using the \verb|fix ti/spring| command \cite{Freitas2016} as implemented in \textsc{lammps}.
For liquid systems, the \verb|pair_style ufm| command \cite{PaulaLeite2016} within \textsc{lammps} is used to model the interatomic interactions in the Uhlenbeck-Ford model, for which the reference-free energy is calculated using the splines provided in Ref. \onlinecite{PaulaLeite2016}. 
In the case of algorithm \ref{algo:step2}, it is necessary to ensure that the system remains in its initial state (solid or liquid) and does not undergo a phase transformation as the Hamiltonian is scaled. 
To this end, Steinhardt's parameters \cite{Steinhardt1983} as implemented in the \textsc{pyscal} code \cite{Menon2019} are used to detect the amount of solid or liquid particles in a given system.

As evident from the sample input configuration, \textsc{calphy} provides an automated way for free energy calculations with minimal user input. 
As classical MD simulations are utilised in \textsc{calphy}, only the vibrational contributions to the free energy are considered.  

The \textsc{calphy} library is available in a public repository \cite{Repository2021}, along with a collection of examples, including those presented in this work.

\section{Conclusion} \label{sec:conclusion}

We implement four different algorithms for the automated calculation of free energies. 
The algorithms can be applied in different scenarios, such as calculating free energies at constant temperature and pressure, calculating the free energy over a given temperature range, and the direct calculation of coexistence lines. 
We demonstrated the efficiency, accuracy, and user-friendliness of our implementation by calculating the complete pressure-temperature phase diagrams of Ti and Si. 
The algorithms can be also be employed for multi-component systems. 
Additionally, we present an algorithm for alchemical changes and upsampling which can efficiently calculate free energies over a given concentration range. 

To facilitate the use of these algorithms, we provide a Python library and command-line program \textsc{calphy}. 
With the help of the framework provided, complex properties such as melting temperature or phase transition temperatures can be easily calculated.

\begin{acknowledgements}

SM, YL and RD acknowledge financial support by the German Research Foundation (DFG) through project number 405621217.
JR acknowledges financial support by the German Research Foundation (DFG) through the DFG Heisenberg Programme project 428315600. The authors
acknowledge computing time by the Center for Interface-Dominated High Performance Materials (ZGH, Ruhr-Universit\"{a}t Bochum).

\end{acknowledgements}

\appendix

\section{Reference free energy \label{app:Hi}}

\subsection{Free energy of solids}
\label{app:Hi:solid}

The Frenkel-Ladd path \cite{Frenkel1984} is commonly used to calculate the Helmholtz free energy in solids from the Einstein crystal. The Hamiltonian of the reference state is given by
\begin{equation}
  H_E = \sum_{i=1}^N \bigg [ \frac{\pmb{p}_i^2}{2m_i} +
    \frac{1}{2} m_i \omega_i^2 (\Delta \pmb{r}_i )^2 \bigg ] \,,
\end{equation}
where $m_i$ is the mass, $\omega_i$ the oscillator frequency and $\Delta \pmb{r}_i$ is the vector of  particle $i$ from its equilibrium position. The Helmholtz free energy is given by
\begin{equation}
  F_E(N,V,T) = 3k_\mathrm{B}T \sum_i \ln \bigg( \frac{\hbar \omega_i}{k_\mathrm{B} T} \bigg) \,.
\end{equation}
The spring constants $k_i = m_i\omega_i^2$ needs to be estimated such that the vibrational frequencies are as close as possible to the solid of interest. 
A common approach~\cite{Freitas2016} is to estimate $k_i$ from the mean-square displacement, $\langle (\Delta \pmb{r}_i)^2 \rangle$,  of the atoms
\begin{equation} \label{eq:spring}
  \frac{1}{2} k_i \langle (\Delta \pmb{r}_i)^2 \rangle = \frac{3}{2} k_\mathrm{B} T \,.
\end{equation}

In addition, a correction due to the fixed centre of mass \cite{Polson2000} needs to be added to the free energy;

\begin{equation}
    \delta F = k_\mathrm{B} T \ln \bigg [ \frac{N}{V} \bigg( \frac{2 \pi k_\mathrm{B} T}{N m \omega^2} \bigg)^{\frac{3}{2}} \bigg ]
\end{equation}

\subsection{Free energy of liquids}
\label{app:Hi:liquid}
The choice of a reference system for a liquid is more complicated than for a solid. A typical reference system is the ideal gas. However, a direct switching path between a liquid and ideal gas can cross the liquid-vapour coexistence line, leading to hysteresis \cite{Abramo2015}. 

The Uhlenbeck-Ford (UF) \cite{PaulaLeite2016} model can be used as a reference system \cite{PaulaLeite2019}. 
The UF model is a purely repulsive pair potential with a single parameter. 
The interaction decays quickly and smoothly providing an advantage over the Lennard Jones potential in terms of truncation or long-range corrections. Furthermore, the UF model only has a stable liquid phase, preventing hysteresis associated with phase transformations. The UF model is given by
\begin{equation}
 H_{UF} = \sum_{i=1}^N  \frac{\pmb{p}_i^2}{2m_i} -  \sum_{i < j}^N p k_\mathrm{B} T \ln (1-\exp{(-(r_{ij}/\sigma)^2)}  )  \,,
\end{equation}
where $r_{ij}$ is the inter-particle distance, $\sigma$ the length scale and $p$ a non-negative scaling factor that controls the strength of the interaction.
The free energy of the UF model is represented as,
\begin{equation} \label{eq:finallqd}
  F_{\mathrm{UF}} = F_{\mathrm{ig}} + F_{\mathrm{UF}}^{(ex)} \,.
\end{equation}
The excess free energy of the UF model is expanded as
\begin{equation}
  F_{UF}^{(ex)}(x, T) = k_\mathrm{B} T \sum_{n=1}^\infty \frac{\tilde{B}_{n+1}(p)}{n} x^n \,,
\end{equation}
with $x \equiv b\rho$ and the number density $\rho$. The reduced virial coefficients $b \equiv (\pi \sigma^2)^{(3/2)}$ and $\tilde{B}_{n+1}(p)$ can be computed exactly. An accurate numerical representation of the free energy using splines is available in literature~\cite{PaulaLeite2016}. The free energy of the ideal gas is given by
\begin{equation}
    F_{\mathrm{ig}}  = N k_\mathrm{B} T \left( \ln \rho - 1 + \sum_n c_n \ln c_n \right) + 3 k_\mathrm{B} T \sum_i \ln \Lambda_i \,.
\end{equation}
The concentration of species $n$ is denoted $c_n$, $\Lambda_i$ is the de~Broglie thermal wavelength
\begin{equation} \label{eq:dbw}
  \Lambda_i = \sqrt{\frac{h^2}{2 \pi k_\mathrm{B} T m_i}} \,.
\end{equation}
where $h$ is the Planck's constant.

\section{Reversible scaling \label{app:revscale}}

Here, we summarise key formulae from Ref. \onlinecite{deKoning1999}.
In reversible scaling, the Hamiltonian is linearly scaled to $\lambda H$. The partition sum $Q$ remains unchanged when $T$ and $\lambda$ are scaled such that $T/\lambda$ is constant. Or, in other words, the change of the Hamiltonian with $\lambda$ has the same effect on the partition function as the scaling of the temperature to $T/\lambda$ while keeping $H$ unchanged. If one further takes into account the temperature scaling in the kinetic energy and the definition of the Helmholtz free energy from the partition function $Q$ as $ F = - k_\mathrm{B} T \ln Q$, one arrives at
\begin{equation}
F(N,V,T/\lambda) =    \frac{3}{2} k_\mathrm{B} T N \frac{\ln \lambda}{\lambda}  + \frac{F(\lambda,N,V,T)}{\lambda} \,.
\end{equation}

where $F(\lambda,N,V,T)$ is the free energy of the scaled Hamiltonian.
Therefore, from computing $F(\lambda,N,V, T)$ the Helmholtz free energy $F(N,V,T/\lambda)$ can be directly obtained, and scaling along $\lambda$ provides the temperature dependence of $F(N,V,T/\lambda)$. For the change of $F$ with $\lambda$ on has
\begin{equation}
\pder{F(\lambda,N,V,T)}{\lambda} = \av{U} \,,
\end{equation}
and
\begin{equation} \label{eq:wa1}
    \Delta F = \int_{1}^{\lambda_f} \av{U} \, d \lambda \, \equiv W_{rev},
\end{equation}
with $T = T_i$ and $\lambda_f = T_i/T_f$.

For the Gibbs energy at a given pressure $P$, a scaling to $\lambda H$ changes the partition function in the isobaric ensemble in the same way as a scaling of temperature and pressure to $T/\lambda$ and $P/\lambda$, therefore
\begin{equation}
G(N,P/\lambda,T/\lambda) =    \frac{3}{2} k_\mathrm{B} T N \frac{\ln \lambda}{\lambda}  + \frac{G(\lambda,N,P, T)}{\lambda} \,. \label{eq:Gscaling}
\end{equation}
This identity may be exploited to evaluate the Gibbs free energy along different pressure paths by assuming that the pressure changes with $\lambda$, $ P = P(\lambda)$. Then
\begin{equation}
\pder{G(\lambda,N,P(\lambda),T)}{\lambda} = \av{U} + \frac{d P(\lambda)}{d \lambda} \av{V} \,,\label{eq:Gpath}
\end{equation}
and
\begin{equation} \label{eq:wa2}
\Delta G = \int_{1}^{\lambda_f} \av{U} + \frac{d P(\lambda)}{d \lambda} \av{V} \, d \lambda \equiv W_{rev}
\end{equation}
Three different pressure paths will be illustrated in the following.

\subsection{Constant pressure}

For computing the temperature dependence of the Gibbs free energy at constant pressure, in Eq.~(\ref{eq:Gscaling}) we change $P$ to $\lambda P$ on both sides so that it becomes
\begin{equation}
G(N,P,T/\lambda) =    \frac{3}{2} k_\mathrm{B} T N \frac{\ln \lambda}{\lambda}  + \frac{G(\lambda,N,\lambda P, T)}{\lambda} \,. \label{eq:Gscaling_const}
\end{equation}
Therefore, for constant pressure simulations the pressure in the scaled Gibbs free energy $G(\lambda,N,\lambda P, T)$ needs to increase linearly with $\lambda$. Eq.~(\ref{eq:Gpath}) reads
\begin{equation}
\pder{G}{\lambda} = \av{U} + P \av{V} \,,\label{eq:Gpath_const}
\end{equation}
and
\begin{equation}
\Delta G = \int_{1}^{\lambda_f} \av{U} + P \av{V} \, d \lambda \,.
\end{equation}

\subsection{Pressure as a function of temperature}

We would like to compute the Gibbs free energy along a given $P$-$T$ path. As the variation of the temperature is achieved through scaling with $\lambda$ from a reference temperature $T$, we use $T/\lambda$ to indicate the varying temperature. Then from Eq.~(\ref{eq:Gscaling}) we have
\begin{equation}
G(N,P(T/\lambda),T/\lambda) =    \frac{3}{2} k_\mathrm{B} T N \frac{\ln \lambda}{\lambda}  + \frac{G(\lambda,N,P_{RS}, T)}{\lambda} \,, \label{eq:Gscaling_PT}
\end{equation}
with the scaled pressure
\begin{equation}
P_{RS} = \lambda P (T/\lambda) \,, \label{eq:PRS}
\end{equation}
and Eq.~(\ref{eq:Gpath}) reads
\begin{equation}
\pder{G}{\lambda} = \av{U} +  \frac{d P_{RS}(\lambda)}{d \lambda}  \av{V} \,,\label{eq:Gpath_PT}
\end{equation}
with
\begin{equation}
\frac{d P_{RS}(\lambda)}{d \lambda} =  P (T/\lambda) - (T/\lambda) \frac{d P(T/\lambda)}{d (T/\lambda)} \,. 
\end{equation}

\subsection{$P$-$T$ coexistence}

The $P$-$T$ scaling may be used for tracking the coexistence boundary between two phases \cite{deKoning2001}. We take $P(T/\lambda)$ as the line along which the Gibbs energy of two phases $\alpha$ and $\beta$ are identical,
\begin{equation}
G_{\alpha}(N,P(T/\lambda),T/\lambda) = G_{\beta}(N,P(T/\lambda),T/\lambda) \,.
\end{equation}
This means that
\begin{equation}
G_{\alpha}(\lambda,N,P_{RS},T) = G_{\beta}(\lambda,N,P_{RS},T) \,.
\end{equation}
As we track the coexistence, a change of $\lambda$ must maintain the condition
\begin{equation}
\pder{{G}_{\alpha}}{\lambda} = \pder{{G}_{\beta}}{\lambda} \,,
\end{equation}
which from Eq.~(\ref{eq:Gpath_PT}) implies
\begin{equation}
\av{U}_{\alpha} +  \frac{d P_{RS}(\lambda)}{d \lambda}  \av{V}_{\alpha} = \av{U}_{\beta} +  \frac{d P_{RS}(\lambda)}{d \lambda}  \av{V}_{\beta} \,,
\end{equation}
and provides a condition for the $P$-$T$ coexistence path,
\begin{equation} \label{Eq:adcc1}
\frac{d P_{RS}(\lambda)}{d \lambda} = \frac{\av{U}_{\alpha} - \av{U}_{\beta}}{\av{V}_{\alpha} - \av{V}_{\beta}} \,,
\end{equation}
which is the Clausius-Clapeyron equation. From this, the pressure change along the coexistence line is obtained,
\begin{equation}
\Delta P = \int_1^{\lambda_f} \frac{ \av{U}_{\alpha} - \av{U}_{\beta} }{ \av{V}_{\alpha} - \av{V}_{\beta} } \, d \lambda \,.
\end{equation}
To make contact to non-equilibrium thermodynamics, the scaling parameter $\lambda$ is varied with time and thermodynamic expectation values are replaced by instantaneous values,
\begin{equation}
\frac{d P_{RS}}{dt} = \frac{d\lambda}{dt} \frac{{U}_{\alpha}(t) - {U}_{\beta}(t)}{{V}_{\alpha}(t) - {V}_{\beta}(t)} \,,
\end{equation}
and the pressure difference is estimated as
\begin{equation}
\Delta P = P_{RS}(\lambda(t_f)) - P_{RS}(\lambda(t_i)) = \int_{t_i}^{t_f} dt \frac{d\lambda}{dt} \frac{{U}_{\alpha}(t) - {U}_{\beta}(t)}{{V}_{\alpha}(t) - {V}_{\beta}(t)} \,.
\end{equation}
This defines the coexistence line $P_{RS}(\lambda)$ in the scaled system. The coexistence pressure in the unscaled system at temperature $T_f$ may be obtained from Eq.~(\ref{eq:PRS}) at  $\lambda(t_f) = T_i/T_f$ by using $\lambda(t_i) = 1$ and $P_{RS}(1) = P_i$ as
\begin{equation}
P_f = P (T_f) = (T_f/T_i) P_{RS}(T_i/T_f) =  (T_f/T_i) (\Delta P + P_i)\,.
\end{equation}

\subsection{Kinetic energy contribution to the free energy} \label{app:mass}

For the purpose of switching the chemistry along an integration path in algorithm \ref{algo:alchemical}, the mass $m$ of the required atoms also needs to change along the path. 
We assume that the mass does not change along the path, and add the kinetic energy contribution to the free energy.

The Hamiltonian of a system of $N$ particles is:

\begin{equation}
    H = \sum_{i=1}^N \frac{\pmb{p}_i^2}{2m} + U(\pmb{r}_1, \pmb{r}_2 ... \pmb{r}_N),
\end{equation}

where $\pmb{p}_i$ and $\pmb{r}_i$ are the momenta and position of particle $i$ and $U$ is the potential energy. The corresponding Helmholtz free energy is,

\begin{equation}
    F(T) = -k_\mathrm{B} T \ln \int d\pmb{r} \exp{( -U/k_\mathrm{B} T )} + 3 k_\mathrm{B} T \sum_{i=1}^N \ln{\Lambda_i(T)}
\end{equation}

$\Lambda(T)$ is the de Broglie wavelength given by Eq.~(\ref{eq:dbw})

Upon a change of mass from $m_i^{(i)}$ to $m_i^{(f)}$ from the initial to final state, the corresponding de Broglie wavelength also changes, which leads to,

\begin{equation}
    \Delta F (T) = 3 k_\mathrm{B}T \sum_{i=1}^N \ln{ \bigg (\frac{ \Lambda_i^{(i)}(T) }{ \Lambda_i^{(f)}(T)} \bigg )} = \frac{3}{2} k_\mathrm{B} T \sum_{i=1}^N \ln{\bigg(\frac{m_i^{(i)}}{m_i^{(f)}}\bigg)}
\end{equation}

\subsection{Switching function} \label{app:switching}

The choice of the functional form of the switching function $\lambda(t)$ in algorithm \ref{algo:step1} affects the energy dissipation during the switching process \cite{deKoning1996}. For algorithm \ref{algo:step1} in solids, the \verb|fix ti/spring| command in \textsc{lammps} \cite{Freitas2016} allows for two functional forms: (i) a linear function $\lambda(t) = t/t_{sw}$ and (ii) a function of the form:

\begin{equation}
    \lambda(t) = {\tau}^5 (70 {\tau}^4-
    315 {\tau}^3+
    540 {\tau}^2-
    420 {\tau}+126)
\end{equation}

where $\tau$ is $t/t_{sw}$. This function has vanishing slopes at the ends of the switching process, and is shown to reduce the energy dissipation \cite{deKoning1996}. Similar to solids, we implement both functional forms for liquids, and provide the option to choose either of the functions. For the other algorithms, we employ a linear function for $\lambda(t)$.

\providecommand{\noopsort}[1]{}\providecommand{\singleletter}[1]{#1}%

\end{document}